\documentclass[sigconf]{acmart}
\usepackage{amsmath}
\usepackage{natbib}
\usepackage{amsfonts}
\usepackage{subcaption}
\usepackage{hhline}
\usepackage{hyperref}
\usepackage{enumerate}
\usepackage{multirow}
\usepackage{array}
\usepackage[para]{footmisc}
\usepackage{bbm}
\usepackage{adjustbox}
\usepackage[algo2e,ruled,vlined]{algorithm2e}
\usepackage{booktabs} 
\usepackage{array}
\usepackage[table]{xcolor}
\usepackage[show]{chato-notes}

\def\tokscore{\mathcal{G}}

\usepackage{tikz}
\newcommand*\circled[1]{\tikz[baseline=(char.base)]{
            \node[shape=circle,draw,inner sep=1pt] (char) {#1};}}

\newcommand{\pageenlarge}[1]{\enlargethispage{#1\baselineskip}}

\copyrightyear{2026}
\acmYear{2026}
\setcopyright{cc}
\setcctype{by}
\acmConference[SIGIR '26]{Proceedings of the 49th International ACM SIGIR Conference on Research and Development in Information Retrieval}{July 20--24, 2026}{Melbourne, VIC, Australia}
\acmBooktitle{Proceedings of the 49th International ACM SIGIR Conference on Research and Development in Information Retrieval (SIGIR '26), July 20--24, 2026, Melbourne, VIC, Australia}
\acmDOI{10.1145/3805712.3809690}
\acmISBN{979-8-4007-2599-9/2026/07}



\begin{document}

\title[PLAID-PRF]{PLAID-PRF: Pseudo-Relevance Feedback \\ with Centroid-like Tokens in PLAID}

\author{Xiao Wang}
\affiliation{%
  \institution{University of International Business and Economics}
  \city{Beijing}
  \country{China}}
\orcid{0000-0002-5151-2773}
\email{xiao.wang@uibe.edu.cn}

\author{Sean MacAvaney}
\affiliation{%
  \institution{University of Glasgow}
  \city{Glasgow}
  \country{United Kingdom}}
\orcid{0000-0002-8914-2659}
\email{sean.macavaney@glasgow.ac.uk}

\author{Craig Macdonald}
\affiliation{%
  \institution{University of Glasgow}
  \city{Glasgow}
  \country{United Kingdom}}
\orcid{0000-0003-3143-279X}
\email{craig.macdonald@glasgow.ac.uk}

\def\name{PLAID-PRF}

\newcommand{\xiao}[1]{#1}
\newcommand{\cm}[1]{#1}
\newcommand{\discuss}[1]{#1}
\newcommand{\sm}[1]{#1}
\newcommand{\smb}[1]{#1}
\newcommand{\smc}[1]{#1}
\newcommand{\smd}[1]{#1}
\newcommand{\x}[1]{#1}
\newcommand{\cmA}[1]{#1}
\newcommand{\xw}[1]{#1}
\newcommand{\xf}[1]{#1}

\newcommand{\crc}[1]{#1}
\newcommand{\crcBeir}[1]{#1}
\newcommand{\crcIo}[1]{#1}
\newcommand{\crcf}[1]{#1}

\begin{abstract}
\looseness -1 Multi-\xw{vector} dense retrieval models, such as ColBERT, achieve strong \crcIo{retrieval} effectiveness by \sm{\x{modelling}} fine-grained token-level interactions between queries and documents.
\smd{Methods such as PLAID use centroid-based quantisation of each token's \smd{vector} to reduce the index size and speed up retrieval while maintaining strong effectiveness.} In this work, we introduce \name{}, a method that performs Pseudo-Relevance Feedback (PRF) over PLAID to reformulate ColBERT's query \smb{vectors}
based on the top-retrieved results. In contrast with prior methods that perform PRF on multi-\xw{\xw{vector}} retrieval models, \name{} keeps computational costs low by leveraging the internal PLAID centroid \smd{vectors}, treating them similarly to tokens in traditional PRF methods. The method selects a small \smb{and} diverse set of high-utility expansion \smd{vectors} and \xiao{appends} them to the original query, rerunning PLAID to refine both candidate generation and final scoring. Extensive experiments on the standard \crcIo{in-domain MSMARCO and four out-of-domain BEIR} benchmarks show that PLAID-PRF consistently improves retrieval effectiveness over various baselines. \x{In particular, PLAID-PRF improves over PLAID by up to 4.3\% nDCG@10 and 7.3\% MRR@10, \smb{while introducing substantially less computation overhead than prior PRF methods}.} The results demonstrate that \smd{our proposed} centroid-aware PRF \smd{method} offers an effective and lightweight mechanism to improve \crcIo{the quality of top-ranked retrieved results}. \crc{Overall, this work enables effective and efficient feedback-aware late-interaction retrieval without expensive \crcf{query-time} document-token \crcf{clustering}}.

\end{abstract}

\maketitle

\section{Introduction}
\looseness -1 Dense retrieval models \sm{encode} different \sm{granularities of} signals \sm{into dense \crcf{(embedding)} \smd{vectors}} depending on their \sm{architecture. A primary way of categorising these models is based on what each \smd{vector} represents:} single-\xw{vector} models \sm{encode one \smd{vector} per retrieval unit (usually passages), while multi-\xw{vector} models encode one \smd{vector} per sub-retrieval unit (usually tokens).}
\sm{Compared to single-\xw{vector} models, multi-\xw{vector} models are capable of performing fine-grained semantic matching over individual tokens---a capability that is lost during the pooling stage of single-\xw{vector} models.}
\sm{For instance, the multi-\xw{vector} model} ColBERT~\cite{khattab2020colbert} and \sm{its derivatives (e.g.,~\cite{santhanam2021colbertv2,wang2023smp,chen-etal-2024-m3,lee2023rethinking,GTE-ModernColBERT,DBLP:conf/ecir/MacAvaneyMT25})} \xiao{perform}
{\em late interaction} \sm{to estimate query-document relevance: the tokens of a query and document are independently encoded using a heavy-weight language model}, and \sm{relevance is estimated} using \sm{light-weight} token-level similarity aggregation (e.g., \crc{MaxSim}). \sm{This process acts like a simple attention mechanism that allows the relevance score to depend only on the parts of each document that are most related to the query, which is a capability missing from single-\xw{vector} dense models because all tokens are pooled into one.}

\begin{figure}[t!]
    \centering
    \includegraphics[width=.9\linewidth]{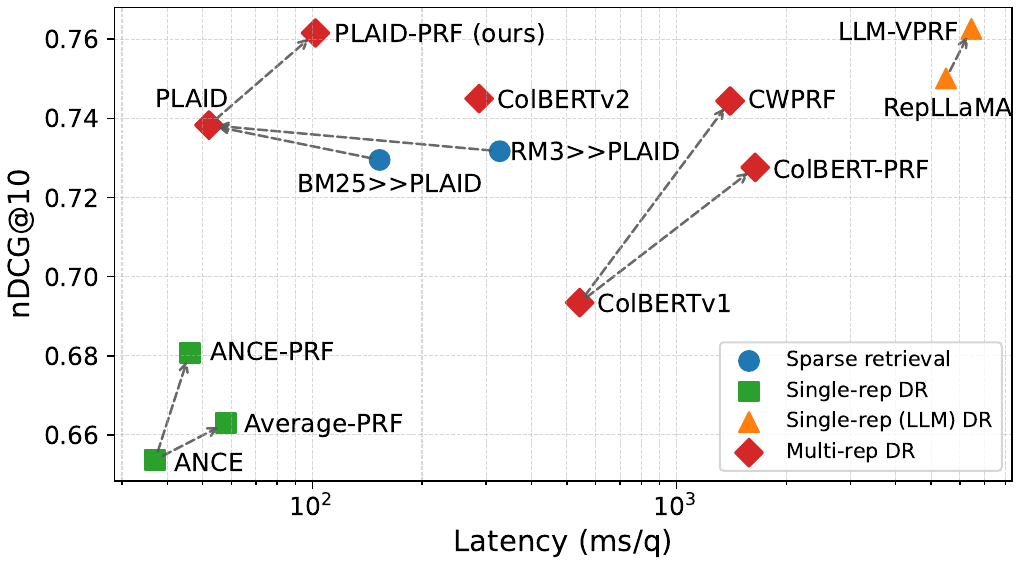}\vspace{-1em}
    \caption{Effectiveness vs. efficiency trade-off on TREC Deep Learning 2019. The y-axis reports nDCG@10, and the x-axis reports query latency (ms/q) on a log scale.}\vspace{-1.5\baselineskip}
    \label{fig:tradeoff}
\end{figure}

\sm{
The main drawback of multi-\xw{vector} models is that they \xf{incur}
substantial storage and memory overheads 
due to having many \smd{vectors} per document (\smd{usually one per token}). \smb{ColBERTv2}~\cite{santhanam2021colbertv2} addressed these drawbacks by clustering and compressing \smd{vectors} during indexing. Specifically, each \smd{token's vector is compressed into} the ID ({\em code}) of its nearest centroid together with a \xiao{quantised} residual, yielding $20$-$32$ bytes per \xw{vector} (depending on the settings), down from $256$ bytes without compression. Later, PLAID~\cite{santhanam2022plaid} leveraged these compressed \xw{vector}s to further reduce approximate retrieval costs by progressively filtering candidate documents through increasingly precise estimations of the model's relevance scores.
}

\sm{
\looseness -1 An orthogonal approach to improve retrieval effectiveness is Pseudo-Relevance Feedback (PRF). PRF typically \xf{consists of}
three phases: (1) initial retrieval for a query\crc{;} (2) reformulation of the query's
representation %
\xf{based on the top-ranked results\crc{;}} and (3) execution of this \xf{refined query representation}~\cite{croft1979using}. 
\xf{PRF \crcIo{approaches have} historically relied on lexical evidence from the corpus, such as word co-occurrences~\cite{lavrenko2001relevance,abdul2004umass}; \crcIo{More} recently, \crcIo{PRF approaches have} evolved to incorporate semantic signals, for instance by leveraging \smd{dense semantic vectors}~\cite{yu2021improving, li2021pseudo} and large language models~\cite{li2025vprf}, to refine the query representation for a second retrieval pass.}
PRF methods have also been successfully designed for multi-\xw{vector} retrieval models, most notably ColBERT-PRF~\cite{wang2022colbert} and CWPRF~\cite{wang2023effective}. Although highly effective, these methods \xf{require computationally expensive clustering or additional model inference}
to select \xf{useful expansion}
\smd{vectors} to add to the query, which substantially increases response times. \xf{In the standard three-stage PRF process,}
reducing the cost of the reformulation step is therefore crucial for the overall efficiency of the method, \crcIo{and its deployability in practical settings.}}

The key insight of this work is that we can leverage the existing \xf{indexing-time}
clusters that PLAID uses for index compression and retrieval to reduce the cost of \xf{PRF-based query reformulation.}
To this end, we introduce \name{}, a training-free PRF framework that directly uses PLAID's \smb{centroid} \xw{vectors}. More specifically, \smd{(1)}~we mine informative \smb{centroids} from the initial top-retrieved set; 
\smd{(2)}~we estimate their usefulness using classical term-weighting models (e.g., TF-IDF)\xf{, computed from {\em centroid} occurrence statistics;} 
and \smd{(3)}~we expand the original query with \smb{a small set of informative and diverse} \smd{vectors from pseudo-relevant documents} and \smb{execute the new query with PLAID}. We find that \crcIo{our proposed \name{} framework}, \cmA{the first PRF work \smb{specifically for} PLAID,} is both inexpensive and effective, setting a new Pareto-optimal operational point as shown in Figure~\ref{fig:tradeoff}. \x{In particular, \name{} improves over PLAID up to +4.3\% \crcIo{in terms of} nDCG@10 on DL'19, and on DL-HARD, it yields a +7.3\% gain in MRR@10, while retaining low-latency retrieval (102 ms/query).}

\section{Related Work}

Pseudo-relevance feedback (PRF) is a long-standing \sm{family of methods which leverage signals from the top-scoring documents from retrieval to inform the rest of the process~\cite{manning2009introduction}.} \sm{PRF methods typically\footnote{A separate class of PRF methods, Adaptive Retrieval, perform only a single stage of retrieval and do not reformulate the query. Instead, they use pseudo-relevance signals to select which documents from the corpus to score\xiao{~\cite{macavaney2022adaptive,rathee2025quam,DBLP:conf/sigir/RatheeVMA25}}. These methods are beyond the scope of this work.} perform automatic query reformulation in three phases: (1) retrieval over the corpus in an initial retrieval stage using the original query\crc{;} (2) reformulation of the query representation %
based on the results from the first stage (e.g., by re-weighting and expanding terms in the query)\crc{;} and (3) executing the reformulated query for the final results.}
\xiao{PRF techniques can be organised into two families: \emph{Discriminative PRF} (Section~\ref{ssec:discriminative}) and \emph{Generative PRF} (Section~\ref{ssec:generative}) techniques. Discriminative PRF methods select and \smb{apply weights to} the information already present in the feedback \smb{documents} (e.g., the expansion terms chosen in the RM3~\cite{abdul2004umass} technique), and \smb{interpolate} them with the original query. \smb{In contrast}, generative PRF methods produce additional text or \xw{vector}s, often generated by 
\smb{Large Language Models (LLMs)}, to augment the original query \smb{with terms that are not necessarily in the feedback documents}. }

\enlargethispage{1\baselineskip}
\subsection{Discriminative PRF}\label{ssec:discriminative}
\looseness -1 \noindent \textbf{Classical sparse PRF. }
Classical lexical-based PRF methods are largely grounded in probabilistic and heuristic techniques. Notable algorithms \smb{include the} \xiao{RM1~\cite{lavrenko2001relevance} and RM3~\cite{abdul2004umass}} \smb{models.} RM1 forms the foundation by estimating a feedback language model directly from the top-k documents to identify likely expansion terms. Building upon RM1, RM3
linearly \crc{interpolates} the RM1 model with the original query, which \smb{helps mitigate}
query drift.
Other influential classical techniques include the Divergence From Randomness (DFR) framework, which includes models like Bo1~\cite{amati2002probabilistic} that weight terms by the divergence of their distribution in the top-k documents from the corpus distribution. 
Methods based on TF-IDF weighting also form a classical basis for selecting salient terms.
\x{These approaches remain strong sparse baselines and are often used as first-stage candidate generators for neural rerankers (e.g., BM25+RM3 followed by BERT/ColBERT reranking)~\cite{wang2022colbert,gao2021complement,haouari2020bigir}.}
However, classical methods operate over bag-of-words evidence and cannot exploit rich contextual information 
in feedback \cmA{documents} (e.g., disambiguating senses of polysemous words). This limitation motivated neural PRF for lexical pipelines, where contextual encoders help select more informative expansions.

\looseness -1 \noindent \textbf{Neural PRF for sparse retrieval. }
Neural PRF methods \smb{leverage} distributed representations to capture semantic relationships. \crc{Early \crcf{embedding-based}
query expansion methods also explored the use of distributed representations in pseudo-relevance feedback settings \cite{zamani2016embedding}.}
Later, 
BERT-QE~\cite{zheng2020bert}  \crc{and}
CEQE~\cite{naseri2021ceqe} \crc{employed} BERT-derived contextual \smb{\xw{embedding}s} to select more semantically-aware expansion terms. While these methods \cmA{improve effectiveness}, they still operate within a lexical framework, leaving a semantic gap between the selected terms and the underlying relevance signal. 
\xiao{Indeed, because they still expand queries in the lexical space, their improvements are ultimately bounded by term selection. This motivates dense PRF methods that perform feedback directly in the embedding space.}

\noindent \textbf{Dense PRF. }
\smb{To overcome limitations of the lexical space}, dense PRF \smb{expands the query using \smd{dense semantic vectors}}.
In single \smb{\xw{vector}} retrieval,  ANCE-PRF~\cite{yu2021improving} \xiao{re-encodes the query together with the PRF documents to produce a refined query \smd{vector}.}
\xiao{\crc{Vector}-PRF~\cite{li2021pseudo} refines the dense query \smd{vector} by aggregating the \smd{vectors} of the top-ranked pseudo-relevant documents (e.g., via averaging or Rocchio-style interpolation).}
\cmA{In contrast, ColBERT-PRF~\cite{wang2022colbert} is designed for multi-\smb{\xw{vector}} settings:} a first pass retrieves with late interaction; feedback token \smd{vectors} are clustered in embedding space \cm{using KMeans; representative centroids} are selected and weighted; a small set of these \smd{vectors are} added to the query before re-execution. Follow-ups such as CWPRF~\cite{wang2023effective} learn contrastive weights to select useful expansion \smd{vectors}.
\cm{By providing additional query \smd{vectors} that are contextualised within the feedback documents, these techniques provide improved semantic matching -- allowing to retrieve further relevant documents even if they are lexically disparate from the feedback documents. However, while ColBERT-PRF and CWPRF can effectively expand the query, they are costly: ColBERT-PRF relies on an expensive clustering of the (large) token \smd{vectors} of pseudo-relevant documents, while CWPRF (like ANCE-PRF) invokes a new transformer model to compute the feedback \smd{vectors}. Instead, we seek an inexpensive PRF stage that can operate effectively within the quantised \smb{\xw{vector}} space used by ColBERTv2/PLAID.}

\subsection{Generative PRF}\label{ssec:generative}
\looseness -1 Generative PRF methods focus on \x{using a generative model
to produce expansion content conditioned on the query and/or feedback, rather than estimating term importance via explicit reweighting.}
These approaches often leverage generative models, including \smb{Large Language Models} (LLMs).
Broadly, they include query-side generation (e.g., GenPRF/GenQR~\cite{wang2023genqr} and LLM-based rewrites) 
or a set of explanatory paragraphs (as in GRF~\cite{iain2023grf} and LLM-VPRF~\cite{li2025vprf}) that are embedded or filtered to guide the second retrieval pass.
\xw{HyDE~\cite{gao2023precise} is a closely-related generation approach that synthesises a hypothetical passage to guide retrieval based on the user's query; however it is not PRF since it does not use pseudo-relevant feedback documents.}
While these methods demonstrate a remarkable capacity for semantic capture and can handle complex queries, the generated content can be verbose or non-specific, so effectiveness on standard retrieval benchmarks is mixed, especially when the first-stage retriever is already strong and in-domain.
Generative PRF is therefore promising 
—especially under domain shift—but \xiao{it} tends to increase inference cost and reduce \xiao{control over the generated expansions, while expanding queries with more diverse terms and paraphrases.}
\x{In this work, we focus on lightweight PRF for quantised late-interaction retrieval, where expansions are identified directly from indexed representations without extra generative inference; generative approaches are orthogonal and are left for future work.}

\section{Preliminaries}\label{sec:preliminary}

\begin{table}[t]
\centering
\small
\setlength{\tabcolsep}{6pt}
\caption{Summary of notation.}
\resizebox{85mm}{!}{
\begin{tabular}{@{}>{$}l<{$} p{0.74\linewidth}@{}} %
\toprule
\multicolumn{2}{@{}l@{}}{\textbf{Notation for ColBERT / ColBERTv2 / PLAID}}\\
\midrule
\phi_{q_i},\phi_{d_j} & \smd{Vector} of the $i$-th query token $q_i$ and $j$-th document token $d_j$ (L2-normalised). \\
\tilde{\phi_{d_j}} & Approximate (reconstructed) \smd{vectors} of the $j$-th document token $d_j$.\\
Q=[\phi_{q_1},\dots,\phi_{q_{n_q}}] & Query representation as a sequence of token \smd{vectors}; length $n_q$. \\
D=[\phi_{d_1},\dots,\phi_{d_{n_d}}] & Document token sequence after ColBERTv1. \\
Score(Q,D) & Late-interaction score calculated using Equation~\eqref{equ:maxsim}.\\
\Psi=\{\mu_c\}_{c=1}^{C} & ColBERTv2 codebook (centroid \smd{vectors}), $C$ is the number of clusters. \\
c_j\in\{1,\dots,C\} & Centroid (code) ID assigned to document token $d_j$. \\
\tilde{r_{j}} & The quantised residual \xw{vector} of document token $d_j$. \\
\midrule
\multicolumn{2}{@{}l@{}}{\textbf{Notation for PLAID-PRF (our work)}}\\
\midrule
\mathcal{P}_{\mathrm{prf}} & PRF set: top \smb{documents} retrieved by a short first pass (stage-1 with the original query). \\
\Phi=\{\tilde{\phi}_{d}\}_{d\in \mathcal{P}_{\mathrm{prf}}} & PRF token \smd{vectors} gathered from $\mathcal{P}_{\mathrm{prf}}$. \\
\mathbf{c}=\{c(\tilde{\phi}_{d})\} & The centroid codes of tokens in $\Phi$ (one code per token). \\
f_p & Number of PRF \smb{documents}. \\
f_e & Number of expansion \smd{vectors} selected. \\
\mathrm{tf}(c) & Frequency of centroid code $c$ among PRF tokens $\Phi$. \\
\mathrm{df}(c) & Global document frequency for code $c$ (precomputed). \\
\tokscore  & Weights of feedback tokens.\\
E \subset \Phi  & Selected expansion \smd{vectors} ($|E| = f_e$).\\%=[\mathbf{e}_1,\dots,\mathbf{e}_{f_e}] %
\beta & Scaling of expansion \smd{vectors} when \crc{appended} to the query ($0<\beta\le 1$). \\
Q_{\mathrm{new}} & Expanded query: $Q_{\mathrm{new}}=[\,Q\oplus \beta\mathbf{e}_1\oplus \cdots \oplus   \beta\mathbf{e}_m\,]$. \\

\lambda & Diversity weight in MMR selection. \\
\bottomrule
\end{tabular}}
\label{tab:notation}
\vspace{-.5em}
\end{table}

\begin{figure}[t!]
\vspace{-.5em}
\centering
\begin{subfigure}[b]{42mm}\centering
\includegraphics[width=42mm]{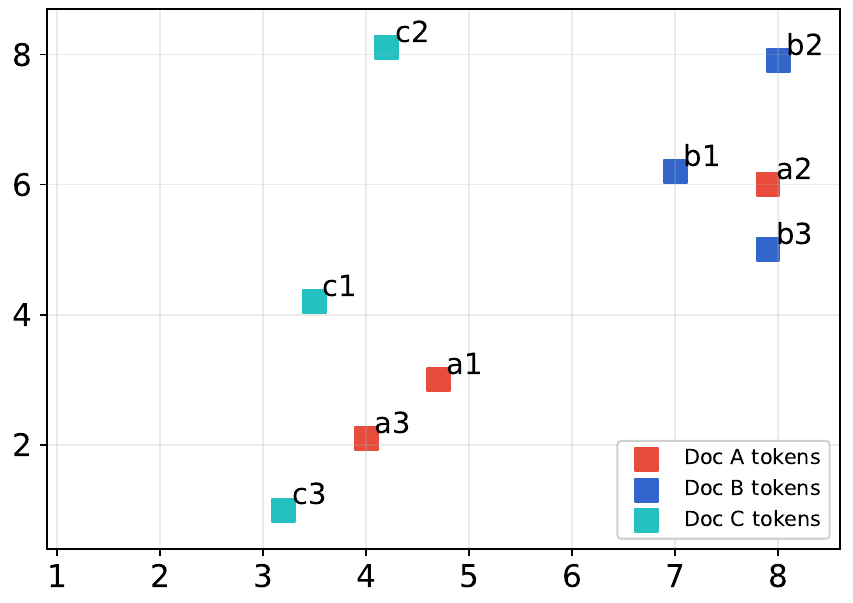}
\caption{ColBERTv1 index stores all \smd{raw} token \smd{vectors}.}\label{fig:colbert}
\end{subfigure}
\begin{subfigure}[b]{42mm}\centering
\includegraphics[width=42mm]{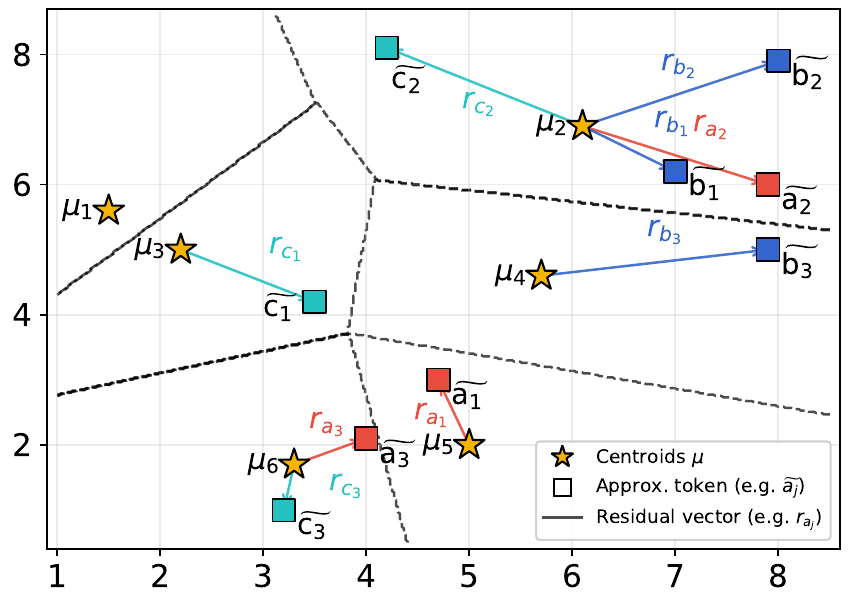}
\caption{ColBERTv2 \& PLAID store \smd{token} centroid ID + residuals}\label{fig:plaid}
\end{subfigure}
\caption{Indexing in ColBERTv1 and ColBERTv2/PLAID.}
\label{fig:index_comp}%
\vspace{-1.5em}
\end{figure}

In this section, we describe the background knowledge of late interaction and explain the ideas that \xiao{motivate}
our method \xiao{using notation summarised in Table~\ref{tab:notation}}. We start \xiao{with}
the late-interaction mechanism introduced in ColBERT~\cite{khattab2020colbert}. We then describe how ColBERTv2~\cite{santhanam2021colbertv2} compresses the document tokens while keeping the same retrieval procedure. After that, we outline the PLAID~\cite{santhanam2022plaid} retrieval pipeline. 

\noindent \textbf{Late interaction in ColBERT.}
ColBERT~\cite{khattab2020colbert} is a dense retrieval model that scores based on the token-level \smd{vectors}. More formally, given a query $Q$ and a \xiao{document $D$,}\footnote{\smb{For simplicity, we refer to all retrieval units (documents, passages, etc.) as documents.}}
ColBERT encodes them into token embeddings\xiao{:} ${Q} = [\phi_{q_1},..., \phi_{q_{|n_q|}} ] \in \mathbb{R}^{n_q \times d}$ and ${D} = [\phi_{d_1},..., \phi_{d_{|n_d|}}] \in \mathbb{R}^{n_d \times d}$, respectively. 

At retrieval time, ColBERT does not pool these \xw{embedding}s into a single \smb{\xw{vector}}. Instead, it \smb{scores using} a late-interaction operator that matches each \smd{vector in the query to the closest vector in the document}:
\begin{equation}\label{equ:maxsim}
\operatorname{Score}(Q, D)=\sum_{i=1}^{n_q} \max _{1 \leq j \leq n_d} Sim(\phi_{q_i}, \phi_{d_j}),
\end{equation}

\looseness -1 This architecture \x{allows fine-grained token-level matching}
\smb{and scores the document based on the \x{tokens that are most semantically}
similar to the query tokens}. In addition, since the \smd{document} token \smd{vectors} \smd{do not depend on the query, they} can be computed offline. 

\noindent \textbf{ColBERTv2 residual representation.} %
ColBERTv2~\cite{santhanam2021colbertv2} retains the late–interaction scoring of ColBERT but compresses document \smd{vectors} to make indexing and search efficient. 
ColBERTv2 learns a global \emph{codebook} of centroid \smd{vectors}
$\Psi=[\mu_c]_{c=1}^{C}$, with each $\mu_c\in\mathbb{R}^{d}$.
Here, $c$ denotes the ID of each centroid, and \(C\) is the number of centroids in the codebook and \(d\) is the \smd{vector} dimension. %
For a document token $d_j$, the indexer stores its nearest–centroid ID
$c_j\in\{1,\ldots,C\}$ together with a quantised residual, $\tilde r_j\in\mathbb{R}^{d}$, \x{that encodes the difference of the \smd{vector} from its centroid ($\phi_{d_j}-\mu_{c_j}$).}
At search time, ColBERTv2 reconstructs a close approximation of the original
token \smd{vector} by adding the centroid back to the residual representation:
\begin{equation}
\label{equ:colbertv2-decompress}
\tilde{\phi}_{d_j}\;\approx\; \mu_{c_j}\;+\;\tilde r_j.
\end{equation}
\looseness -1 The codebook \(\Psi\) is stored once with the index; individual tokens carry only the integer
\(c_j\) and their compressed residual.

\pageenlarge{1}
\looseness -1 The residual remains $d$-dimensional, but each dimension is quantised to a few bits (e.g., 1--2), then decompressed at query time for approximate reconstruction.
\xiao{The centroid-oriented \cmA{nature of the resulting index} enables efficient candidate generation, followed by residual-aware refinement to preserve accuracy.
This residual representation-based index structure also yields a substantial reduction in index size: on MS MARCO passages, ColBERTv2 achieves a 6--10$\times$ compression ratio, reducing the index from 154~GiB to 16~GiB (1 bit per dimension) or 25~GiB (2 \crc{bits} per dimension).}
As we argue in Section~\ref{sec:plaid-prf}, 
\smd{this approximate representation of the document vectors} has two key implications for our work. 
\cm{First, every document token is associated to a given centroid ID \(c_j\) -- we observe that occurrences of {\em centroids} are countable, and hence we can measure their informativeness.}
Second, because centroids form the backbone of the index, reasoning at the centroid level is fast and stable: code–level signals are cheap to compute and align with how the system stores and accesses \xw{vector}s. \cm{This suggests that use of (pre-computed) centroids within PRF may bring significant efficiency benefits over online computation of clusters used in ColBERT-PRF.}

\noindent \smc{\textbf{PLAID search pipeline.} To further increase the retrieval efficiency of ColBERTv2, 
PLAID~\cite{santhanam2022plaid} searches over encoded ColBERTv2 \smd{vectors} in a multi-stage cascading pipeline. First, it identifies a large candidate set by scoring the centroid \smd{vectors} against the query (i.e., performing $\Psi \cdot Q^{\top}$) and mapping the top-scoring centroids for each query \smd{vector} back to the documents that contain them using an inverted index. It then produces increasingly fine-grained approximations of each document's score, truncating the ranked list after each approximation. The final stage performs final MaxSim scoring over the reconstructed document \smd{vectors}. The process is controlled by several parameters, which are studied in detail in~\cite{macavaney2024reproducibility}.}

The three architectures above\xiao{, namely ColBERTv1 \& v2 and PLAID,} point to a clear path towards \name{}. \cm{Firstly, the summative nature of MaxSim allows additional query \smd{vectors} obtained by PRF to improve effectiveness, as demonstrated by ColBERT-PRF \& CWPRF\xiao{.} Secondly, ColBERTv2 provides a useful semantic unit  -- the centroid code -- for identifying possible expansions\xiao{;}} Thirdly, the PLAID pipeline reveals where extra query \smd{vectors} generate the most value: 
by enabling the query to probe more centroids, which can uncover \cm{additional relevant documents}, \cm{while also improving \crcIo{the precision of the top-ranked results.}}

\section{PLAID-PRF}\label{sec:plaid-prf}
\begin{figure}[tb]
    \centering
 \includegraphics[width=.9\linewidth]{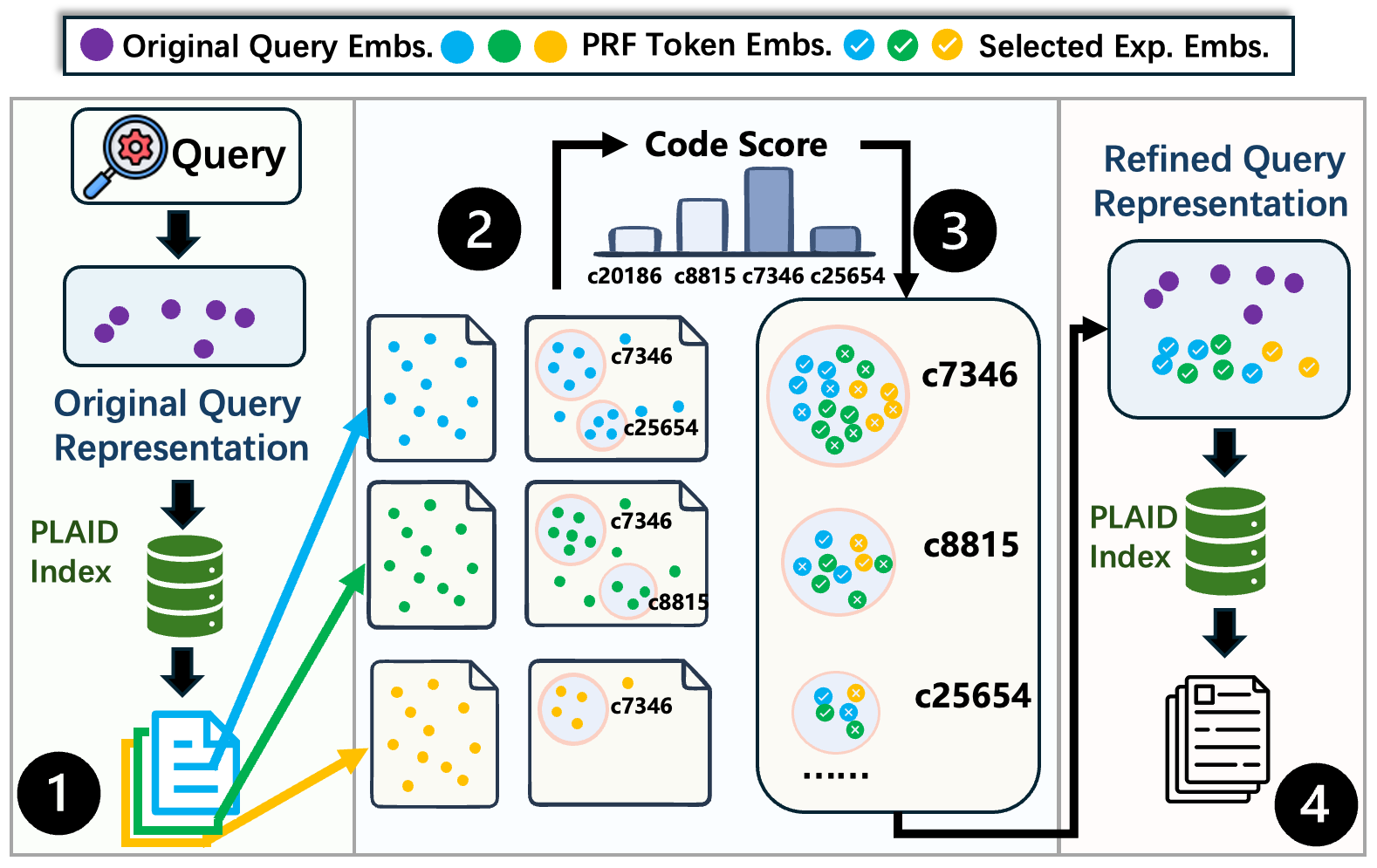}\vspace{-1em}
    \caption{Overview of PLAID-PRF. An initial PLAID retrieval returns pseudo-relevant documents, where the feedback token \smd{vectors} are stored in the PLAID index as centroid codes with quantised residuals. PLAID-PRF groups the tokens by their centroid code (e.g. c7346/c8815) and calculates the code-level score. PLAID-PRF ranks centroids by their scores and assigns higher usefulness to tokens originating from higher-scored centroids. Finally, a small set of highly-useful expansion \smd{vectors} is selected and expanded to form a refined query representation for final retrieval. 
    }
    \label{fig:plaid_prf_framework}
    \vspace{-\baselineskip}
\end{figure}

In this section, we introduce~\name{}\xiao{, which is illustrated in Figure~\ref{fig:plaid_prf_framework}}.
\name{} augments PLAID's late-interaction retrieval with a small set of weighted expansion \smd{vectors} derived from the initial top-retrieved documents. \cm{ We describe \name{} in four stages, as highlighted in the figure: \circled{1} an initial retrieval; \circled{2} estimating the informativeness of centroids occurring in the retrieved documents; \circled{3} identify a set of expansion \smd{vectors} associated to the informative centroids to form the refined query representation; and \circled{4} a final second-pass retrieval. Table~\ref{tab:notation} provides an overview of our used notation, while Sections~\ref{ssec:stage1}-\ref{ssec:stage4} provide an overview of the four stages of \name{}.}

\subsection{Stage 1: Initial Retrieval}\label{ssec:stage1}

\looseness -1 Given a tokenised query, the encoder produces a sequence of
\smd{vectors} $Q = \big[\,\phi_{q_1},\ldots,\phi_{q_{n_q}}\,\big]$, 
we perform \xiao{an initial}
PLAID search to obtain a \textit{feedback set} of the top–$f_p$ documents,
\(\mathcal{P_{\mathrm{prf}}}=\{p_1,\ldots,p_{f_p}\}\). 
\cmA{Further, by concatenating all $\mathcal{P_{\mathrm{prf}}}$ documents, we can obtain a sequence of {\em feedback tokens}.} In \xiao{a} PLAID/ColBERTv2 index, each token's \smd{vector} is associated with a centroid code \(c_j\) (from the indexing-time codebook) along with a quantised residual \xiao{$r_j$} used for reconstruction. \cmA{These are used for retrieval, and available for PRF.}

\subsection{Stage 2: Usefulness Scoring}\label{ssec:stage2}%

\xf{In PRF, we typically assume that the top-ranked feedback documents are relevant. However, they likely also contain redundant and noisy tokens, so naively appending all feedback token \smd{vectors} can drift the query representation away from its original intent. Therefore, within the feedback token set, we estimate the usefulness of each token and incorporate the \smd{vectors} of the most discriminative tokens into the query representation.}

\looseness -1 An informative token (and its corresponding centroid code) should be \emph{discriminative} in two senses: it should occur frequently among the top feedback documents for this query, reflecting \emph{topical relevance}, yet appear rarely across the collection overall, reflecting \emph{specificity}. In ColBERT-PRF~\cite{wang2022colbert}, \xiao{the expansion mechanism implicitly balances topicality and specificity: topical relevance is captured by query-conditioned cluster centroids formed over pseudo-relevant token \smd{vectors},} while specificity is measured by ranking clusters using the IDF of the nearest token to each centroid.

\looseness -1 \cm{We observe that tokens \xf{sharing} a centroid tend to be projected in a similar region of the \smd{vector} space.}
\xf{This implies that, instead of performing expensive retrieval-time clustering (as in ColBERT-PRF), we can reason in terms of ColBERTv2 centroid occurrences, which are already stored in the index. A centroid that occurs frequently in the feedback documents can be considered useful for query reformulation. We therefore compute usefulness weights at the centroid level and propagate these weights to the tokens assigned to each centroid.}

\cm{\cmA{Indeed}, instead of counting the specificity of tokens directly,} the centroid codes \cm{enable us to approximate} the discriminative properties of each feedback token. In other words, we can compute classical IR statistics (TF, IDF, etc.) \cm{upon the occurrences of centroids}. 
\footnote{\crc{Indeed, prior work has also shown that corpus-statistics-based reweighting \crcIo{of centroid \smd{vectors}} can be useful for retrieval using embedding-based representations~\cite{arora2017simple,galke2017word}.}} %
\xf{Unlike}
the classic PRF term weighting method over lexical terms, \name{} \xf{focuses}
on centroid IDs: every occurrence of the centroid code $c$ in the PRF set can be quantified as $\mathrm{tf}(c)$, and its global rarity can be measured using a centroid-level IDF.  The latter is easily computed offline using the PLAID index.

Then, we \xiao{map}
these weights back to the tokens based on the positionally aligned assignments between centroid codes and tokens. \cmA{We provide further details of centroid weighting and token-level usefulness scores below.}

\paragraph{Centroid weighting}

\looseness -1 Let $\Phi$ denote the sequence of all feedback token \smd{vectors} in $\mathcal{P_{\mathrm{prf}}}$, i.e. $\Phi=[\tilde{\phi}_{d_j}]_{j=1}^{|\Phi|}$ for each feedback token $j$, along with corresponding centroid codes \([c_j]_{j=1}^{|\Phi|}\). 
For each centroid $c$, we can count $tf(c) = \sum_{j=1}^{|\Phi|} \mathbbm{1}(c_j = c),$  i.e., the number of feedback tokens assigned to centroid $c$ for the current query. Similarly,  we can \xf{pre-}compute 
document frequency $\mathrm{df}(c)$ \xf{at indexing time}, i.e. how many documents \crc{in the index} contain at least one token assigned to centroid $c$. 
This allows \xf{us} to calculate a TF-IDF weight for the occurrence of each centroid within the feedback set, as: $w_{\mathrm{tfidf}}(c) \;=\; \mathrm{tf}(c)\cdot \frac{N+1}{\mathrm{df}+1}$. Intuitively, TF-IDF gives high weights to centroids that are common in the feedback (indicating topical relevance) but rare in the corpus\footnote{We did observe that the $\mathrm{df}(\cdot)$ distribution exhibits approximately power-law–like behaviour over a limited range of frequencies ($10^3\ldots 10^4$).} (indicating specificity) — and therefore capturing the discriminative nature of each cluster. 
\cm{As we show later, this 
repurposing of existing term weighting models applied to centroid occurrences works pleasantly well.} Indeed, \xiao{in our experiments, we investigate various classical feedback usefulness weighting \cm{models} in Section~\ref{ssec:ablations}.}

\paragraph{Token-level \xiao{usefulness}}

Given the per-centroid usefulness weights $w(c_j)$ computed above, each feedback token is assigned a final score, $\tokscore \in \mathbbm{R}^{|\Phi|}$ using the weight of its corresponding centroid code:
\begin{equation}
    \tokscore = [ w(c_j) ]_{j=1}^{|\Phi|}.
\end{equation}
\looseness -1 $\tokscore$ is used to determine which of the reconstructed feedback \smd{vectors} $\Phi$ to add back into the query representation, as described next.

\subsection{Stage 3: Expansion \smd{Vector} Selection}\label{ssec:stage3}

\looseness -1 Given the (reconstructed) feedback \smd{vectors} $\Phi$ and their token-level usefulness scores $\tokscore$ (Section~\ref{ssec:stage2}), we select a compact set of $f_e$ expansion \smd{vectors} $E \subset \Phi$ to append to the query representation.
A naive Top-$k$ policy picks the $f_e$ tokens with the largest scores in $\tokscore$.
\cmA{However, as centroids can represent more than one feedback token, this policy can result in very similar (i.e. redundant) \smd{vectors} being selected.}
To encourage \cm{a more diverse set of expansion \smd{vectors}}, we adopt greedy maximal marginal relevance (MMR)\xiao{~\cite{carbonell1998mmr,clarke2008mmr,santos2010exploiting}}, which \cmA{\crc{greedily}
selects \smd{vectors}} into $E$ that are highly weighted in $\tokscore$ yet dissimilar to previously selected \smd{vectors} and the original query \smd{vectors}:
\begin{equation}\small
j^{*}=\arg\max_{j\notin E_t}\left[(1-\lambda)\,\tokscore_j-\lambda \max_{v\in E_t\cup Q}\mathrm{Sim}(\tilde{\phi}_{d_j},v)\right],
\end{equation}
\looseness -1 where \x{we select an expansion \smd{vector} indexed by $j^*$ at each iteration step $t$,}
$E_t$ is the current selected set, $Q$ denotes the original query \smd{vectors}, and $\lambda\in[0,1]$ controls the usefulness--diversity trade-off.\footnote{We min-max normalise $\tokscore$, such that they lie in a compatible range with cosine similarity $\mathrm{Sim}()$.} We \x{update $E_{t+1}=E_t \cup\tilde{\phi}_{d_{j^*}}$ and} iterate until $|E_t|=f_e$. \cmA{Notably, the expansion \smd{vectors} $E$ are formed of reconstructed \smd{vectors} $\tilde{\phi}_{d_j}$, rather than centroid \smd{vectors} $\mu_j$ - indeed, as we show later, centroid \smd{vectors} are too coarse for effective query refinement.}

\subsection{Stage 4: Expand the query and rerun PLAID}\label{ssec:stage4}

\xiao{We} append the selected expansion \smd{vectors} $E$ to form \xiao{the refined query representation:}
\begin{equation}
Q_{\text{new}}=\big[\,Q \oplus  \beta \cdot E]
\in\mathbb{R}^{(n_q+f_e)\times d},
\end{equation}
where \(\beta\in(0,1]\) controls the feedback contribution.
We then rerun the PLAID retrieval using the refined query representation.

\subsection{Discussion}

\looseness -1 \smc{Dense PRF methods such as ColBERT-PRF and CWPRF operate directly on the full-precision token \smd{vectors} of ColBERTv1. In contrast, ColBERTv2 avoids the intractable memory requirements of storing such \smd{vectors} by relying on quantisation, which precludes direct access to full-precision token representations. As a result, \name{} is designed for quantised late interaction and operates over reconstructed \smd{vectors}. Specifically, it estimates topical relevance using \emph{PLAID’s centroid codes}, treating centroid IDs as semantic “terms”, and subsequently maps the resulting weights back to the corresponding reconstructed token \smd{vectors}. This design allows PLAID-PRF to reuse PLAID’s compressed, centroid-based index structure and multi-stage probing without retraining the retriever, while still supporting term-based weighting schemes and MMR-style selection over reconstructed expansion \xw{vector}s. Central to this design are three modelling choices: leveraging PLAID’s indexing-time clusters rather than computing retrieval-time clusters as in ColBERT-PRF; counting centroid frequencies instead of WordPiece occurrences; and performing expansion over reconstructed \smd{vectors} rather than directly using centroids. To our knowledge, PLAID-PRF is the first PRF framework that is both \emph{codebook-aware} and \emph{quantisation-compatible}, aligning feedback statistics with the structure of a ColBERTv2/PLAID index. Our experiments aim to validate our three modelling choices, as described next.}

\section{Experimental Setup}

Our first experiments focus upon the key \xiao{modelling} decisions in \name{}:

\textbf{RQ1.} \xiao{Is feedback based on indexing-time clustering 
more effective than \xf{feedback based on }PRF-stage clustering}?

\looseness -1 \textbf{RQ2.} Does codebook-based centroid counting yield more effective expansions than using WordPiece token statistics?

\textbf{RQ3.} \cmA{Is \xf{the} use of the reconstructed token \smd{vectors} necessary for effective PRF, or can raw centroid \smd{vectors} be used for expansion?}

\noindent Then, we compare the retrieval performance of \name{} to various existing baselines \smd{and using out-of-domain benchmarks}: 

\textbf{RQ4.} How does PLAID-PRF compare against state-of-the-art retrievers and PRF methods in effectiveness and latency?

\noindent \crc{Furthermore,}
we conduct a \cmA{sensitivity analysis} for \name{}:

\looseness -1 \textbf{RQ5.} How sensitive is \name{} to (i) the expansion settings (number of feedback documents etc.); (ii) choice of centroid weighting model, and (iii) diversity-based expansion selection setting?

\crc{\textbf{RQ6.} Which expansion \smd{vectors} help or hurt PLAID-PRF?}

\crc{\textbf{RQ7.} Does PLAID-PRF generalise to out-of-domain retrieval, as measured on selected BEIR datasets? }

\subsection{Setup}%

\looseness -1 \textbf{Datasets and Metrics.} Experiments are conducted \cmA{in-domain} on MS MARCO passage corpus~\cite{nguyen2016ms}, using the TREC Deep Learning 2019~\cite{craswell2019overview} and 2020~\cite{craswell2020overview}
query sets. The corpus contains 8.8M passages gathered from Bing search results. For TREC DL 2019 and 2020, we use the 43 and 54 topics with multi-grade relevance assessments, respectively. On average, these test collections provide around 95.4 and 66.8 relevance judgements per query for 2019 and 2020 respectively, giving sufficiently deep labels to support the evaluation of pseudo-relevance feedback methods~\cite{wang2022colbert,yu2021improving,li2021pseudo}.\footnote{MS MARCO dev set offers very shallow annotations (on average, 1.1 relevant passages per query); PRF can be unreliable on such sparsely judged test collections~\cite{amati2004query,wang2022improving},
therefore we omit the Dev queries.} Moreover, we also evaluate using DL-HARD~\cite{mackie2021dlhard} query sets, which consist of 50 most challenging DL topics according to the SERP-based criteria. %
\x{We focus on in-domain TREC DL benchmarks to isolate the effect of our codebook-based expansion under a controlled evaluation setting.}
\crc{Moreover, we report results on four representative test collections from BEIR benchmark suite~\cite{thakur2021beir}, namely DBPedia, NFCorpus, TREC-COVID, and Touché-2020. Following prior work~\cite{wang2022improving}, we restrict evaluation to BEIR test collections with a sufficiently large number of relevance judgements per query, as measurements of pseudo-relevance feedback are known to be less reliable on test collections with sparse judgements~\cite{amati2004query}.}

\looseness -1 Following the standard practice of recent dense retrieval works, we report mean reciprocal rank (MRR) and normalised discounted cumulative gain (nDCG) calculated at rank 10 as our primary effectiveness metrics, together with \xiao{R}ecall and mean average precision (MAP) calculated at rank 1000. For binary metrics, namely MRR, MAP, and Recall, we consider labels of two or greater to be relevant~\cite{craswell2019overview}. For efficiency, we report the mean response time (MRT), i.e., the average per-query latency (ms/query) measured with a query batch size of 1. For statistical significance testing, we use the paired t-test with \x{Holm–Bonferroni multiple testing correction ($p < 0.05$)}.
\crc{For the BEIR evaluation, we report nDCG@10 and Recall@1000 to compare both early ranking quality and high-depth retrieval coverage~\cite{thakur2021beir,kamalloo2023resources}.}

\looseness -1 \noindent \textbf{Parameters.} 
For PLAID-PRF, we use the official ColBERTv2 checkpoint,\footnote{https://huggingface.co/colbert-ir/colbertv2.0} trained on the MS MARCO passage ranking \xiao{corpus}~\cite{nguyen2016ms}, and build a PLAID index with residual compression. Each token's \smd{vector} is stored as a centroid identifier plus a \xiao{quantised} residual, using 2-bit residual \xiao{quantisation} (\texttt{nbits}=2). For retrieval, we enable the PLAID engine and adopt the recommended operational point from~\cite{santhanam2022plaid,macavaney2024reproducibility}, corresponding to ($n_{cells}=4$, $t_{cs}=0.40$, $n_{docs}=4096$). %
Based on the parameter study in \xiao{Section~\ref{ssec:ablations}}, \xiao{we tune PLAID-PRF on TREC DL'19 via a grid search over $f_p, f_e, \beta, \lambda$ and fix the parameters for all other test sets. We select the configuration based on nDCG@10 and set}
$f_p=3$ feedback passages, $f_e=14$ expansion \smd{vectors}, $\beta=0.7$ and $\lambda=0.3$. 
\crcBeir{For the BEIR experiments, we keep the structural PLAID-PRF parameters fixed and tune only the feedback scaling factor $\beta$. This is a practice analogous to the interpolation weight commonly used in classical PRF methods such as Rocchio- and RM3-style feedback~\cite{rocchio1971relevance,abdul2004umass}.}
For estimating the usefulness of centroids, in addition to TF-IDF, we also investigate other heuristic weighting methods, including RM1~\cite{lavrenko2001relevance}, RM3~\cite{abdul2004umass} and Bo1~\cite{amati2002probabilistic}, as well as a strawman random weighting of centroids. All experiments of \name{} are conducted using a single
NVIDIA GeForce RTX~3090 GPU. 
\crc{The source code for reproducing PLAID-PRF is available in the PyTerrier\_ColBERT2 repository.\footnote{\url{https://github.com/cmacdonald/pyterrier_colbert2/tree/main}} We also provide a virtual appendix containing the result files to produce the result tables in this paper.\footnote{\url{https://github.com/Xiao0728/PLAID-PRF-VirtualAppendix/tree/master}}}

\subsection{Baselines}\pageenlarge{1}

We compare 
\name{} with the following 4 families of retrieval systems, applied with and without PRF.

\noindent \textbf{Sparse Retrieval.} We compare with the classical sparse  retrieval systems, namely BM25, and with RM3~\cite{abdul2004umass} and PRF technique implemented. We implement these systems using PyTerrier~\cite{macdonald2021pyterrier}.\footnote{https://github.com/terrier-org/pyterrier} Moreover, we also compare both BM25 and BM25 + RM3 systems with PLAID as a reranker, which are implemented following~\cite{macavaney2024reproducibility}.%

\noindent \textbf{Single-\smd{vector} dense retrieval.} We include ANCE~\cite{xiong2020approximate} as a standard dual-encoder dense retriever trained on MS~MARCO, together with two PRF variants: ANCE-PRF~\cite{yu2021improving}, which applies classical term-based PRF on top of ANCE scores, and Average-PRF~\cite{li2021pseudo}, which performs an average operation for the feedback \smd{vectors} in the dense representation space.\footnote{https://github.com/terrierteam/pyterrier\_dr/}

\noindent \textbf{Single-\smd{vector} LLM-based dense retrieval.} We compare with the more-powerful LLM-based dense retrieval approaches and their PRF extensions, namely PromptReps (Qwen 8B)~\cite{zhuang2024promptreps}
and PromptPRF (8B)~\cite{li2025pseudo} \crc{(which augments PromptReps with LLM-generated features)}. Moreover, we \xiao{implement}
\crcf{RepLLaMA}
(7B)~\cite{ma2024repllama} and its LLM-VPRF~\cite{li2025vprf} variant \xiao{with FAISS~\cite{johnson2019billion} HNSW~\cite{malkov2020hnsw} search,}
which applies \xw{vector}-space PRF on top of RepLLaMA.\footnote{As PLAID is a compressed ANN index representation, for fair comparison we similarly apply RepLLaMA using an ANN, specifically using HNSW with settings proposed in~\cite{ladr}.} Together, these baselines represent the state of the art in single-rep LLM-centric dense retrieval and PRF. %

\noindent \textbf{Multi-\smd{vector} dense retrieval.} Finally, we compare with multi-\smd{vector} models. ColBERTv1~\cite{khattab2020colbert} is the original late-interaction retriever operating on token-level representations; Col\-BERT-PRF~\cite{wang2022colbert} and CWPRF~\cite{wang2023effective} extend it with PRF in the token space.\footnote{ https://github.com/terrierteam/pyterrier$\_$colbert} We also compare with ColBERTv2~\cite{santhanam2021colbertv2} and PLAID~\cite{santhanam2022plaid}.\footnote{https://github.com/seanmacavaney/plaidrepro}

\section{Results and Analysis}\label{sec:res}

\begin{table}[]
    \caption{Effectiveness of PLAID-PRF Variants.  $\dagger$ \x{indicates a significant improvement over PLAID, while $\ddagger$ indicates the variant is significantly outperformed by PLAID-PRF}
    (paired test, $p<0.05)$. The highest value in each column is boldfaced.}
    \vspace{-.5\baselineskip}
    \centering
    \begin{adjustbox}{max width=85mm}
    \begin{tabular}{l c cc  cc cc}
    \toprule
    \multirow{2}{*}{Methods} 
    & \multicolumn{2}{c}{TREC 2019} 
    & \multicolumn{2}{c}{TREC 2020} 
    & \multicolumn{2}{c}{DL-HARD}\\
    \cmidrule(lr){2-3}\cmidrule(lr){4-5}\cmidrule(lr){6-7}
    & nDCG@10  &MRR@10 &nDCG@10 &MRR@10 &nDCG@10 &MRR@10  \\
\midrule
PLAID 
 & 0.7383 & 0.8934  &0.7410 &0.8411
 &0.4146 &0.5697
\\
LocalCluster-PRF
& 0.7523%
& 0.8879 &0.7552 &0.8298
&0.4160 &0.5476
  \\
TokenCount-PRF%
& 0.7610%
& 0.9310  &0.7573 &0.8486 
&0.4258 &0.5526
 \\ 
CentroidExp-PRF 
 &0.7502%
 & 0.8820  &\bf{0.7581} &0.8294 
 &0.4002$\ddagger$ &0.5046$\ddagger$\\

PLAID-PRF   

&\bf{0.7700}$\dagger$ &\bf{0.9287}  &{0.7435} &\bf{0.8488}
&\bf{0.4351}$\dagger$ &\bf{0.6110}\\
\bottomrule
    \end{tabular}
    \end{adjustbox}
    \label{tab:variants}
\end{table}

\begin{figure}[tb]
    \centering
    \includegraphics[width=\linewidth]{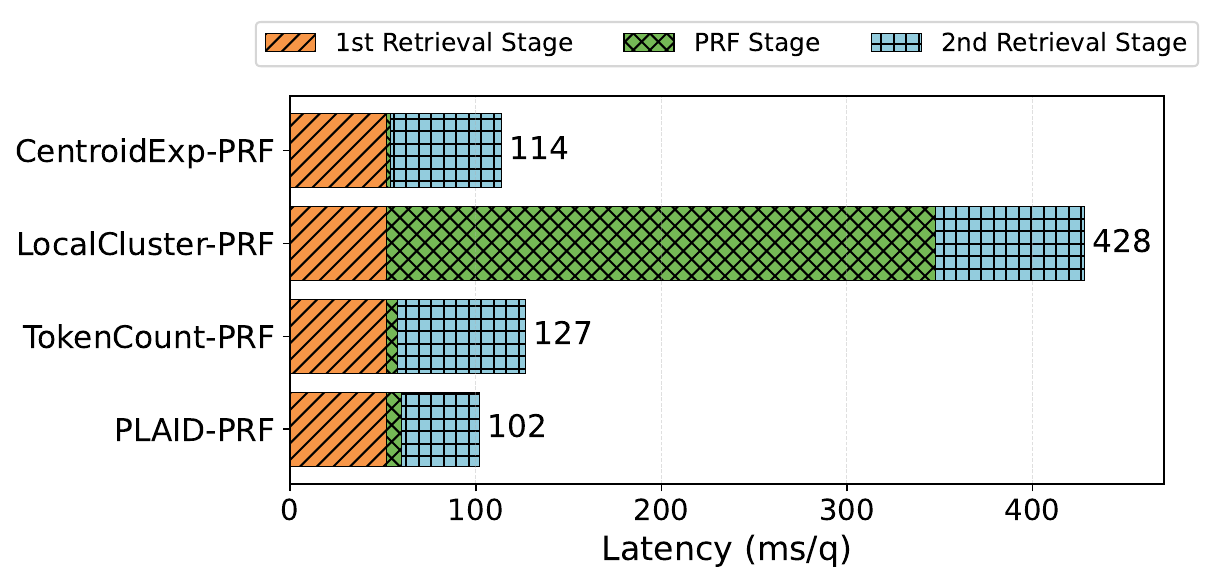}\vspace{-1.3em}
    \caption{Latency Breakdown of PLAID PRF variants.}
    \label{fig:variants} 
\end{figure}

\pageenlarge{1}
\looseness -1 \cmA{We firstly address RQs 1-3 -- concerned with design choices in \name{} -- in Sections~\ref{ssec:rq2} - \ref{ssec:rq3}. In addressing these RQs, we make use \xf{of} Table~\ref{tab:variants} (effectiveness of variants) and Figure~\ref{fig:variants} (efficiency of variants).}
Section~\ref{ssec:rq4} compares \name{} to state-of-the-art (cf. RQ4), while Section~\ref{ssec:ablations} reports the sensitivity analysis (\xf{cf. }RQ5). 
\crc{Furthermore, Section~\ref{ssec:qualitative} presents the qualitative analysis of~PLAID-PRF (cf. RQ6) and Section~\ref{ssec:ood} \crcf{examines} the out-of-domain generalisation of~PLAID-PRF (cf. RQ7).}

\subsection{RQ1: Indexing-time clustering vs. PRF-stage clustering}\label{ssec:rq2}

\enlargethispage{.5\baselineskip}
\looseness -2 \noindent \textbf{Experimental Setup.}
We compare PRF based on \emph{indexing-time clustering} (PLAID-PRF) against a PRF variant based on \emph{PRF-stage clustering} \cmA{(denoted LocalCluster-PRF). Indeed, LocalCluster-PRF is essentially ColBERT-PRF using a KMeans but using the reconstructed token \smd{vectors} rather than exact token \smd{vectors} (which are not stored in a ColBERTv2 index).} 
The two approaches differ only in the clustering source -- the global PLAID codebook with $k=2^{18}$ clusters versus per-query KMeans with $k=24$ -- while keeping \xiao{the 1st and 2nd pass retrieval \crc{configurations} unchanged.}\footnote{Following ColBERT-PRF~\cite{wang2022colbert}, we set $k=24$ for PRF-stage KMeans.}

\looseness -1 \noindent \textbf{Results.}
Table~\ref{tab:variants} shows that PLAID-PRF yields consistently higher early-precision effectiveness than LocalCluster-PRF. 
\x{In particular, \name{} improves over LocalCluster-PRF on all reported metrics across the three query sets, for example, on DL'19, \name{} substantially improves nDCG@10 (0.7523 $\rightarrow$ 0.7700).}
\x{Meanwhile, e}fficiency differences are even larger: Figure~\ref{fig:variants} reports \x{423}
ms/query for PRF-stage clustering, versus 102\,ms/query for \name{} \x{(4.15$\times$ faster),}
with most of the excess latency attributable to the runtime KMeans over feedback token \smd{vectors}.
\x{We postulate that centroid-only expansions are less well aligned with PLAID’s codebook, which reduces selectivity and increases the amount of second-pass refinement and late-interaction scoring.}
\cmA{This shows the advantage of using PLAID's codebook, computed at indexing time, rather than an expensive KMeans clustering  for \x{each}
query at inference time.}
Overall, \x{in response to \textbf{RQ1}, }indexing-time clustering yields a markedly better effectiveness--efficiency trade-off.

\subsection{RQ2: Counting centroids or tokens}\label{ssec:rq1}

\noindent \textbf{Experimental Setup.}
Since the PLAID index stores token representations in compressed form (centroid codes with quantised residuals) and does not retain the original WordPiece IDs, we precompute a mapping from each passage’s token positions to their WordPiece token IDs using the same BERT tokeniser as in PLAID indexing. This yields a strict one-to-one alignment between (i) the token text produced by the tokeniser, (ii) its WordPiece ID, and (iii) its position in the PLAID token sequence, enabling us to recover WordPiece IDs for the pseudo-relevant tokens returned in the first pass. \cmA{We then implement a PRF variant, denoted TokenCount-PRF, that computes  TF–IDF weights at the WordPiece level, rather than the centroid-code level as per \name{}.}

\looseness -1 \noindent \textbf{Results.} Table 2 shows that replacing WordPiece-level counting (TokenCount-PRF) with codebook-aware centroid counting (PLAID-PRF) yields consistently stronger retrieval effectiveness.
This is because centroid counting treats PLAID’s centroid IDs as semantic ``terms'', mapping multiple tokenisation variants and morphological forms to shared semantic units, which reduces redundant expansions and \xf{better aligns the feedback evidence with PLAID’s codebook representation.}
This tends to promote the most relevant documents earlier (higher MRR), whereas TokenCount-PRF can be sensitive to tokenisation and affixes. 
In terms of retrieval latency, \crc{as} shown
in Figure~\ref{fig:variants}, \name{} is also faster than TokenCount-PRF (102 \x{vs.} \x{127}
ms/q), suggesting that code-level counting not only improves ranking quality but also avoids the overhead associated with \xiao{WordPiece}-level token alignment in the retrieval pipeline. 
\x{Overall, in response to \textbf{RQ2}, we find that codebook-based centroid counting yields more effective and more efficient expansions than WordPiece-level statistics.}
\vspace{-.5\baselineskip}

\subsection{RQ3: Reconstructed \smd{vectors} vs. raw centroid \smd{vectors} as expansion units}\label{ssec:rq3}

\noindent \textbf{Experimental Setup.}
We examine whether the choice of expansion units matters when performing PRF in a PLAID setting. Specifically, we compare \name{}, which expands the query with reconstructed token \smd{vectors} (centroid \smd{vectors} plus quantised residual), against a centroid-only variant (denoted CentroidExp-PRF) that uses raw codebook centroids as expansion \smd{vectors}.

\noindent \textbf{Results.}
Table~\ref{tab:variants} shows that CentroidExp-PRF is competitive in nDCG@10, achieving the best DL’20 nDCG@10 (0.7581)—but yields lower early precision than \name{} 
\x{on both \crc{reported} metrics for DL'19 and DL-HARD query sets.} 
This suggests that using raw centroids without residuals yields a coarser expansion signal: it can improve overall top-10 ranking (nDCG@10) yet is less effective at promoting the first relevant item to the very top ranks (MRR@10). In contrast, reconstructed token \smd{vectors} retain centroid+residual information and better match MaxSim
late interaction, yielding stronger MRR. This shows the necessity of using reconstructed \smd{vectors} for effective expansion, answering {\bf RQ3}.
\x{Finally, we note that CentroidExp-PRF shows a comparable end-to-end latency to PLAID-PRF (114 vs. 102 ms/q), but is consistently slower due to a heavier second-pass retrieval stage.}

\begin{table*}[h!]
    \centering  \vspace{-.5\baselineskip}
    \caption{Retrieval effectiveness and efficiency of \name{} and various baselines. A superscript $\dagger$ on a model's score indicates that \name{} significantly outperforms that baseline (paired test, $p<0.05$). The highest value in each column is boldfaced.}%
    \vspace{-.5\baselineskip}
    \begin{adjustbox}{max width=180mm}
\begin{tabular}{l cccc cccc cccc  c}
 \toprule 
\multirow{2}{*}{Methods} & \multicolumn{4}{c}{TREC 2019 (43 queries)} & \multicolumn{4}{c}{TREC 2020 (54 queries)}  & \multicolumn{4}{c}{DL-HARD (50 queries)} &\multirow{2}{*}{Latency (ms)} \\
 \cmidrule(lr){2-5}\cmidrule(lr){6-9}\cmidrule(lr){10-13}
   &MAP  &nDCG@10 &MRR@10 &R@1k %
      &MAP  &nDCG@10 &MRR@10 &R@1k 
      &MAP  &nDCG@10 &MRR@10 &R@1k
      & \\
 \midrule
  \multicolumn{14}{c}{Multiple \smd{Vector} Dense Retrieval}\\
\midrule
ColBERTv1~\cite{khattab2020colbert} 
&0.4318$\dagger$ &0.6934$\dagger$ &0.8527 &0.7892$\dagger$ 
&0.4654$\dagger$ &0.6871$\dagger$ &0.8500 &0.8245$\dagger$ 
&0.2358$\dagger$ &0.3824$\dagger$ &0.5291 &0.7586$\dagger$
&542
\\

ColBERT-PRF~\cite{wang2022colbert}
&0.5295 &0.7283 &0.8508 &0.8746
&0.4904 &0.6937$\dagger$ &0.7684$\dagger$ &0.8914
&0.2306$\dagger$ &0.3683$\dagger$ &0.4546$\dagger$ &0.8021
&1645
\\

CWPRF~\cite{wang2023effective}
&\bf{0.5335} &0.7444 &0.8671 &0.8596 
&0.5145 &0.7246 &0.8421 & 0.8783 
&-&-&-&-
&1403
\\
 \cmidrule(lr){2-14}

ColBERTv2~\cite{santhanam2021colbertv2}
&0.5118 &0.7448$\dagger$ &0.8973 &{0.8940}
&0.5270 &0.7401 &0.8408 &\bf{0.9129}
&0.2727$\dagger$ &0.4144$\dagger$ &0.5720  &\bf{0.8655}
&287\\

PLAID~\cite{santhanam2022plaid} 
&0.5097 & 0.7383$\dagger$ & 0.8934 & 0.8706 
&0.5262 &0.7410 &0.8411 &{0.9044} 
&0.2727$\dagger$ &0.4146$\dagger$ &0.5697 &0.8389 
&52 \\

\midrule

  \rowcolor{gray!20}  %
 \bf{\name{} (Ours)} %
 &0.5282 &\bf{0.7700} &\bf{0.9287} &\bf{0.8967}
 &\bf{0.5236} &\bf{0.7435} &\bf{0.8482} &0.8906
 &\bf{0.2905} &\bf{0.4351} &\bf{0.6110} &0.8485
 &102
 \\
 
 \midrule
  \multicolumn{14}{c}{Sparse Retrieval Systems}\\
\midrule
 BM25 
 & 0.2864$\dagger$ & 0.4795$\dagger$ & 0.6396$\dagger$ &0.7553$\dagger$ 
 & 0.2930$\dagger$  & 0.4936$\dagger$  & 0.6147$\dagger$  &0.8072$\dagger$  
 &0.1471$\dagger$ &0.2743$\dagger$ &0.4151 &0.6690$\dagger$ 
 &80
 \\
 
BM25 + PLAID~\cite{macavaney2024reproducibility}
 &0.4815 &0.7295 &0.8760 &0.7553$\dagger$
 &0.5091 &0.7435 &0.8584 &0.8072$\dagger$  
 &0.2520$\dagger$ &0.4000$\dagger$ &0.5372 &0.6866$\dagger$ 
 &153
 \\

BM25 + RM3
 &0.3192$\dagger$ &0.5247$\dagger$ &0.6462$\dagger$ &0.7859$\dagger$ 
 &0.3161$\dagger$  &0.5107$\dagger$  & 0.6141$\dagger$  &0.8333$\dagger$ 
 &0.1646$\dagger$ &0.2843$\dagger$ &0.4095 &0.7103$\dagger$ 
 &253 \\
 BM25 + RM3 + PLAID
 &0.4942 &0.7423  &0.8798 &0.7859$\dagger$ 
 &0.5091 &0.7401 &0.8693 &0.8333$\dagger$  
 &0.2560$\dagger$ &0.4080 &0.5406 &0.7103 $\dagger$
 &327
 \\

\midrule
\multicolumn{14}{c}{Single \smd{Vector} Dense Retrieval
}\\
\midrule

ANCE~\cite{xiong2020approximate}
&0.3716$\dagger$ &0.6210$\dagger$ &0.7876$\dagger$ &0.7234$\dagger$
&0.3871$\dagger$ &0.6077$\dagger$ &0.7460$\dagger$ &0.7423$\dagger$
&0.1839$\dagger$ &0.2980$\dagger$ &0.4364$\dagger$ &0.6892$\dagger$
&37
\\
ANCE-PRF~\cite{yu2021improving}
&0.4296$\dagger$ &0.6409$\dagger$ &0.7829$\dagger$ &0.7237$\dagger$
&0.4090$\dagger$ &0.6160$\dagger$ &0.7594 &0.7400$\dagger$
&0.1955$\dagger$ &0.2981$\dagger$ &0.4474$\dagger$ &0.6554$\dagger$
&58
\\

Average-PRF~\cite{li2021pseudo} %
&0.4163$\dagger$ &0.6303$\dagger$ &0.7442$\dagger$ &0.7155$\dagger$
&0.4080$\dagger$ &0.6197$\dagger$ &0.7459$\dagger$ &0.7282$\dagger$
&0.1927$\dagger$ &0.2995$\dagger$ &0.4029$\dagger$ &0.6375$\dagger$
&46
\\

\midrule
 \multicolumn{14}{c}{Single \smd{Vector} (LLM) Dense Retrieval}\\
\midrule

PromptReps (8B)~\cite{zhuang2024promptreps} %
&0.3165$\dagger$ &0.5422$\dagger$ &0.6940$\dagger$ &0.7377$\dagger$
&0.2734$\dagger$ &0.4960$\dagger$ &0.6549$\dagger$ &0.7282$\dagger$
&- &- &- &- &- 
\\
PromptPRF (8B)~\cite{li2025pseudo}
&- &0.5941 &- &0.7176
&- &0.5265 &- &0.7040 
&-&-&-&- 
&- \\

RepLLaMA (7B)~\cite{ma2024repllama}
&0.4325$\dagger$ &0.6117$\dagger$ &0.7543$\dagger$ &0.7044$\dagger$ 
&0.4977 &0.7241 &0.8843 &0.8189$\dagger$ 
&0.2628 &0.4132 &0.5343 &0.7496$\dagger$ 
&5496\\

LLM-VPRF~\cite{li2025vprf}
&0.4409$\dagger$ &0.6300$\dagger$ &0.7775$\dagger$ &0.8905$\dagger$
&0.5222 &0.7335 &0.8525 &0.8786$\dagger$
&0.2753 &0.4378 &0.5576 &0.7791
&6454
\\

 \bottomrule
\end{tabular}
\end{adjustbox}
\label{tab:main_res}
\vspace{-0.5\baselineskip}
\end{table*}

\subsection{RQ4: \name{} vs.\ SOTA}\label{ssec:rq4}
\xiao{In this section, we evaluate \name{} wrt.\ state-of-the-art retrievers in terms of both effectiveness and efficiency. 
Table~\ref{tab:main_res} reports the main results across query sets, while Figure~\ref{fig:tradeoff} visualises the retrieval effectiveness-latency trade-off on DL'19. We first discuss effectiveness and then analyse latency.}

\noindent \textbf{Effectiveness of \name{}.}
From Table~\ref{tab:main_res}, we observe that PLAID-PRF delivers the strongest top-ranked effectiveness while retaining the efficiency advantages of PLAID-style approximate late interaction compared to various baseline families.

Firstly, we examine the effectiveness of the proposed PRF mechanism by comparing \name{} \xiao{with}
PLAID. From Table~\ref{tab:main_res}, we find that \name{} yields significant improvements over PLAID. More specifically, it improves nDCG@10 by +4.3\% 
(0.7383$\rightarrow$0.7700) and MRR@10 by +4.0\% (0.8934$\rightarrow$0.9287) on DL'19 and even higher improvements on DL-HARD, with nDCG@10 increasing by +4.9\% (0.4146$\rightarrow$0.4351) and MRR@10 by +7.3\% (0.5697$\rightarrow$0.6110).
These gains in early-precision metrics indicate that PRF expansions do not merely increase recall, but more importantly promote relevant documents to higher ranks.
\xiao{Similarly,} \name{} \xiao{significantly} improves over ColBERTv1 and ColBERTv2 models. 
Moreover, PLAID-PRF also \xiao{exhibits}
large gains over prior token-level PRF methods: on DL’20 it improves ColBERT-PRF by +6.3\% nDCG@10 (0.6993$\rightarrow$0.7435) and +5.5\% MAP (0.4962$\rightarrow$0.5236), and surpasses CWPRF by +2.6\% nDCG@10 (0.7246$\rightarrow$0.7435).

Furthermore, we also compare against strong sparse pipelines and single-\smd{vector} dense models (both conventional and LLM-based).
In particular, within sparse retrieval, PLAID-PRF improves over strong cascades, \xf{which}
employ PLAID as reranker in BM25+PLAID and BM25+RM3+PLAID baselines.
Secondly, compared with single-\smd{vector} dense retrievers (ANCE and its PRF variants), PLAID-PRF is \xiao{significantly}
stronger on both DL’19 and DL’20 (e.g., DL’20 nDCG@10: 0.7435 vs. 0.6077 for ANCE, +22.3\%), consistent with the known advantage of late interaction for fine-grained evidence alignment.
Finally, recent LLM-based retrievers form a strong (often SOTA) baseline family. PLAID-PRF remains highly competitive in this regime, typically outperforming prompt-based dense retrievers and RepLLaMA.
This indicates that an efficient token-level PRF mechanism atop PLAID can rival LLM-based retrieval pipelines without requiring substantially heavier inference and computational resources.

\noindent \textbf{Efficiency of \name{}.} %
From Table~\ref{tab:main_res} and Figure~\ref{fig:tradeoff}, \name{} achieves a strong effectiveness--efficiency balance and lies on the Pareto frontier on DL’19. Figure~\ref{fig:variants} further decomposes the end-to-end latency of \name{} into three stages: the 1st-pass retrieval, the PRF stage, and the 2nd-pass PLAID rerun. Overall, \name{} \xiao{requires}
102 ms/query, where the PRF stage contributes only a small fraction of the total runtime. This is because \name{} computes the expansion usefulness scores through lightweight centroid-level statistics and directly forms expansion \smd{vectors} from the retrieved token representations. Importantly, the refined query representation is constructed in the \smd{vector} space and passed to the second-stage PLAID rerun without re-encoding the query, so the overall cost remains dominated by the two PLAID retrieval passes rather than the feedback computation.

\subsection{RQ5: Sensitivity Analysis}\label{ssec:ablations}

\begin{figure*}[t]
    \centering
    \begin{subfigure}[t]{0.32\textwidth}
        \centering
        \includegraphics[width=\linewidth]{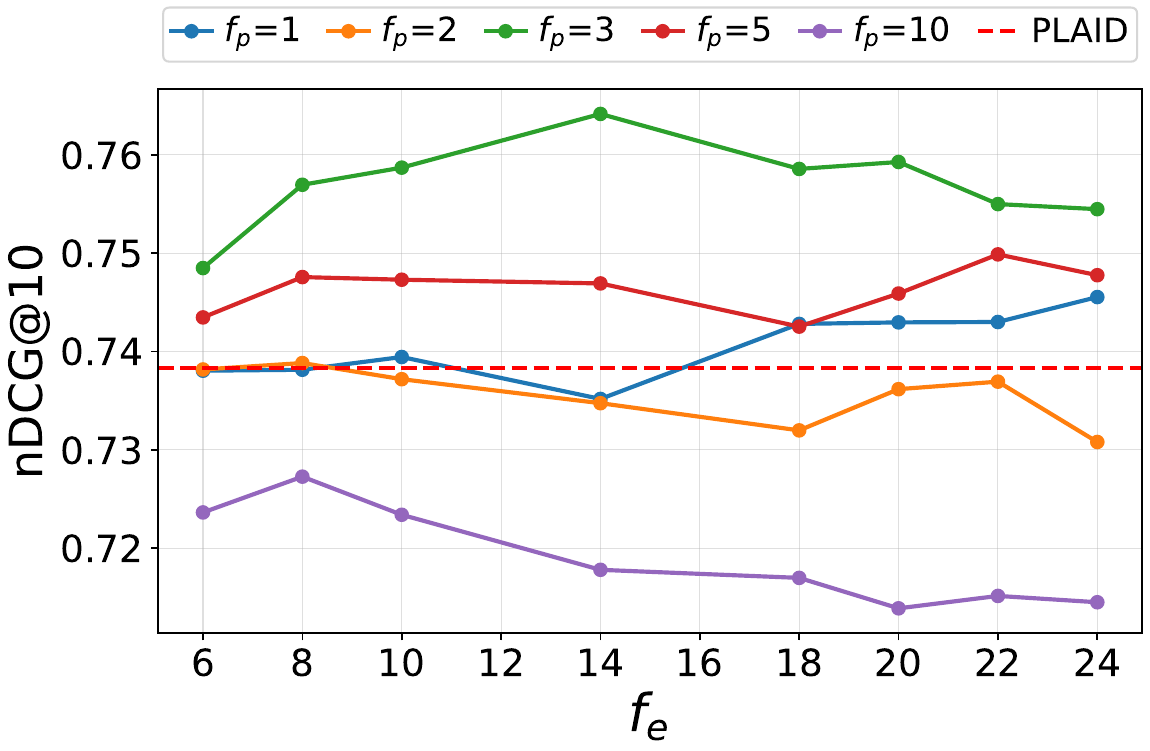}\vspace{-.5\baselineskip}
        \caption{Impact of $f_p$ \& $f_e$}
        \label{fig:para:a}
    \end{subfigure}\hfill
    \begin{subfigure}[t]{0.32\textwidth}
        \centering
        \includegraphics[width=\linewidth]{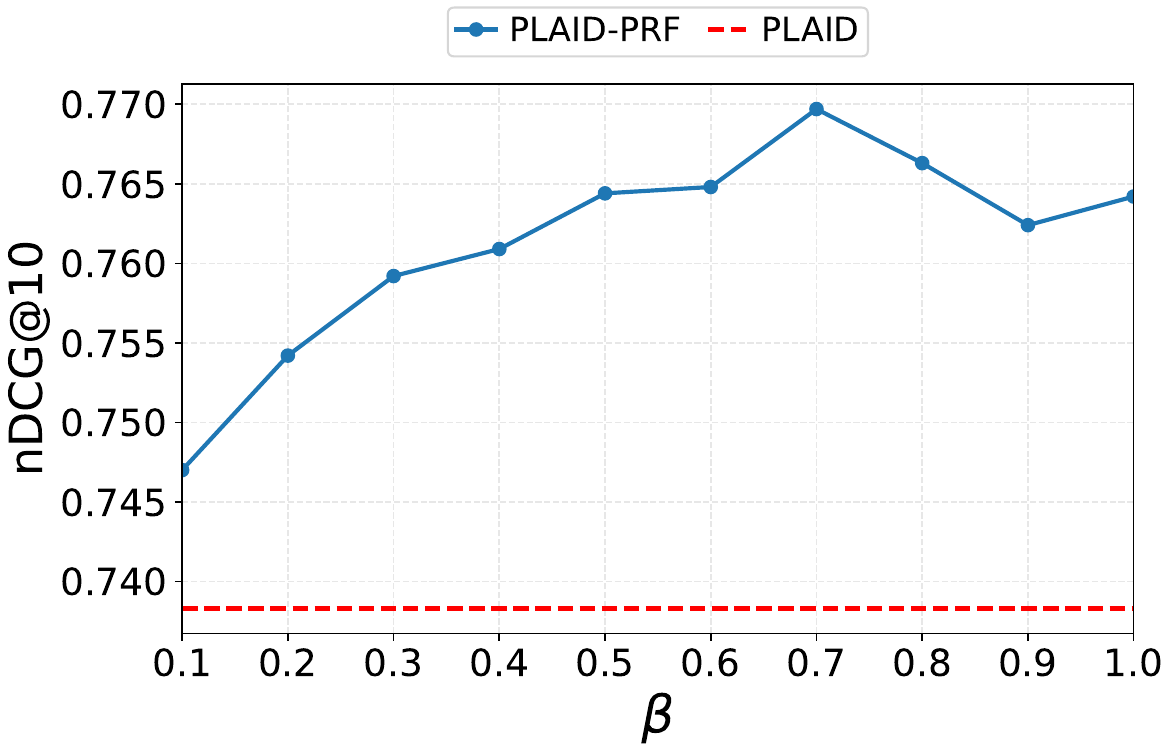}\vspace{-.5\baselineskip}
        \caption{Impact of $\beta$}
        \label{fig:para:b}
    \end{subfigure}\hfill
    \begin{subfigure}[t]{0.32\textwidth}
        \centering
        \includegraphics[width=\linewidth]{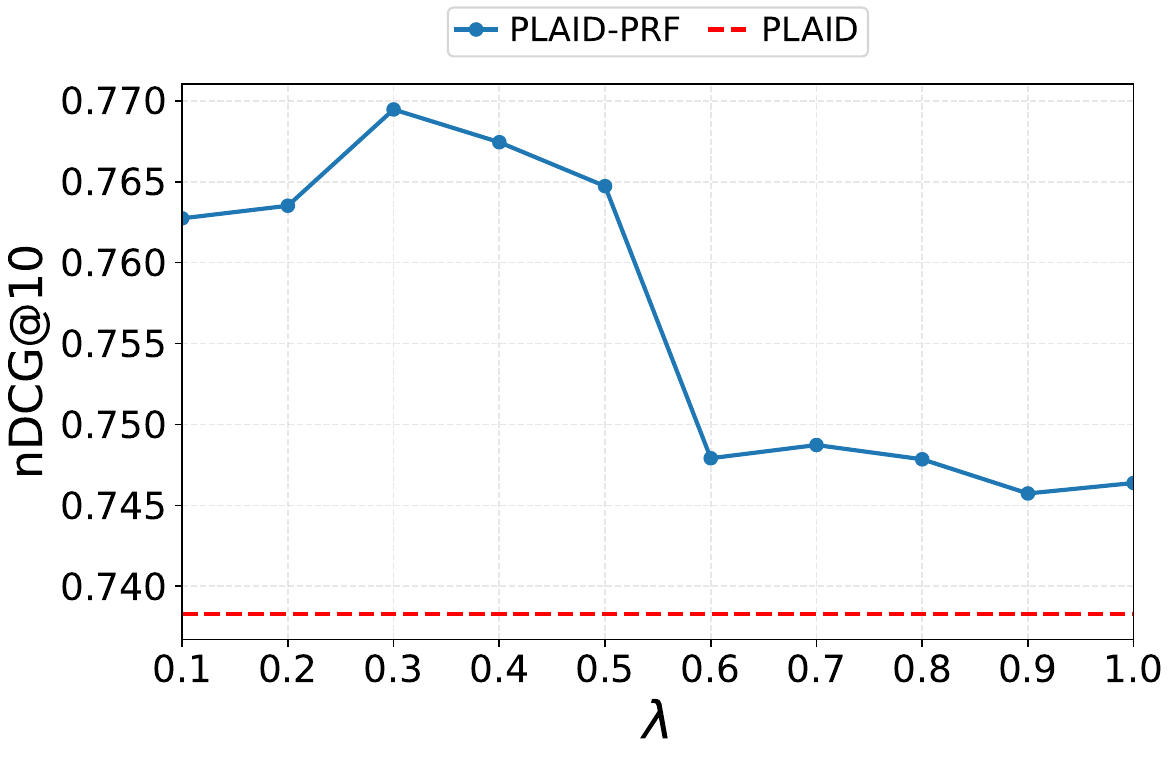}\vspace{-.5\baselineskip}
        \caption{Impact of $\lambda$}
        \label{fig:para:c}
    \end{subfigure}\hfill
    \vspace{-.5\baselineskip}
    \caption{Impact of $f_p$, $f_e$, $\beta$ and $\lambda$ on the Retrieval Effectiveness (nDCG@10 on TREC 2019 queries) of \name{}.}
    \vspace{-.5\baselineskip}
    \label{fig:para}
\end{figure*}

\noindent \textbf{Impact of the PRF parameters.}
Figure~\ref{fig:para} studies four PRF parameters with clear operational meanings: $f_p$ controls how many pseudo-relevant passages contribute feedback evidence, $f_e$ controls how many expansion \smd{vectors} are appended to the query, $\beta$ controls the interpolation strength of the expansion
and 
$\lambda$ controls the relevance–diversity trade-off in MMR selection. Across all settings, many configurations exceed the \xiao{effectiveness of} PLAID baseline (dashed line), showing that PLAID-PRF’s gains are not confined to a single fragile point.

\looseness -2 More specifically, in terms of $f_p$ and $f_e$ (Figure~\ref{fig:para:a}), increasing 
$f_p$ from 1 to 3 improves nDCG@10, as more top-ranked passages provide a broader feedback signal, \xf{however, at larger $f_p$ values (e.g. 10), nDCG@10 degrades due to noisier or less relevant feedback.}
This trend is consistent with \xf{findings in }classical PRF~\cite{abdul2004umass,amati2002probabilistic} and dense-PRF~\cite{yu2021improving, wang2022colbert},
where small feedback depths are often preferred to limit \crc{query}
drift. For $f_e$, performance rises with more expansions but saturates and can regress when $f_e$ becomes large, reflecting diminishing returns and redundancy among expansion \smd{vectors}. The curve peaks around $f_p=3, f_e=14$, which we adopt.
In terms of $\beta$ and $\lambda$ (Figure~\ref{fig:para:b} \& Figure~\ref{fig:para:c}, respectively)\x{, we observe that nDCG@10 increases as $\beta$ rises from 0.1 to 0.7, then slightly decreases as $\beta$ increases further (0.8–1.0).} 
The best DL’19 point occurs near $\beta=0.7$, but $\beta\in[0.4,0.7]$ remains consistently strong, so we treat this as a robust range. 
$\lambda$ peaks at a small value ($\lambda = 0.3$: mild diversity regularisation helps avoid near-duplicate expansions, while larger $\lambda$ over-emphasises diversity and hurts relevance. Overall, these trends motivate our final settings: $f_p = 3, f_e=14, \beta=0.7$ (robustly $\beta\in [0.4,0.7]$), and $\lambda=0.3$.

\begin{table}[tb]
    \centering
    \caption{Impact of the weighting method \& expansion selection method for \name{}. Notations per Table~\ref{tab:variants}}\vspace{-.5\baselineskip}
    \begin{adjustbox}{max width=86mm}
    \begin{tabular}{l|cc cc cc}
    \toprule
    \multirow{2}{*}{Methods} & \multicolumn{2}{c}{TREC 2019} & \multicolumn{2}{c}{TREC 2020} &\multicolumn{2}{c}{DL-HARD}\\
    \cmidrule(lr){2-3}\cmidrule(lr){4-5} \cmidrule(lr){6-7}
     & nDCG@10  &MRR@10 &nDCG@10 &MRR@10 &nDCG@10 &MRR@10  \\

    \midrule
    PLAID  & 0.7383 & 0.8934  &0.7410 &0.8411 &0.4146 &0.5697\\
    \midrule
    \multicolumn{7}{c}{Impact of weighting method}\\
    \midrule
    PLAID-PRF (Rand. W.) 
    &0.7362$\ddagger$
    &0.8632$\ddagger$
    &0.7471 &0.8127
    &0.4179 &0.5319%
    \\
    PLAID-PRF (RM1)     &0.7616 &0.9171  &\bf{0.7561} &0.8482  &\bf{0.4367}$\dagger$ &0.5831\\
    PLAID-PRF (RM3)     &0.7630 &0.9132  &0.7549 &0.8575
    &0.4365$\dagger$ &0.5847\\
    PLAID-PRF (Bo1)     &0.7659 &0.9054  &0.7434 &0.8482  
    &0.4316 &0.5874\\
    PLAID-PRF (TF-IDF)  &\bf{0.7700}$\dagger$ &\bf{0.9287}  &0.7435 &\bf{0.8488}  &0.4351$\dagger$ &\bf{0.6110}\\
    \midrule
    \multicolumn{7}{c}{Impact of expansion selection method}\\
    \midrule
    PLAID-PRF (Rand. Sel.) 
    &0.7413 &0.8740 &\bf{0.7555} &0.8480 &0.4140 &0.5476\\
    PLAID-PRF (Top-k) &0.7594 &0.9132 &{0.7439} &\bf{0.8520}
    &0.4274 &0.5938\\
        
    PLAID-PRF (MMR)   &\bf{0.7700}$\dagger$ &\bf{0.9287} &0.7435 &0.8488
    &\bf{0.4351}$\dagger$ &\bf{0.6174}\\
    
    \bottomrule
    \end{tabular}
    \end{adjustbox}
    \label{tab:weighting_mmr}
    \vspace{-\baselineskip}
\end{table}

\looseness -1
\noindent \textbf{Impact of the weighting method.}
Table~\ref{tab:weighting_mmr} reports ablations on the weighting scheme and the expansion selection strategy in \name{}. 
We find that using random weights 
eliminates the benefits of PRF: effectiveness becomes unstable and early precision 
degrades, indicating that gains come from meaningful token importance estimation rather than simply appending extra \smd{vectors}.
In addition, replacing TF–IDF with alternative weighting methods shows that several choices remain competitive, but TF–IDF provides the most robust gains overall: it achieves the strongest effectiveness on DL’19 (nDCG@10$\simeq$ 0.770 with significant improvements over PLAID) and consistently high MRR across test sets, while also maintaining strong performance on DL-HARD. In contrast, RM-style weighting (RM1/RM3) 
can yield higher top-10 gain on DL’20 (RM1 attains the best nDCG@10), yet is less consistent across collections. \xiao{Bo1 is generally weaker, likely because centroid codes may not exhibit the frequency statistics that Bo1 expects (see Section 4.2).}

\noindent \textbf{Impact of expansion selection method.}
For selecting expansion \smd{vectors}, Table~\ref{tab:weighting_mmr} shows that random selection underperforms both MMR and Top-$k$, highlighting that effective expansion requires principled token choice and redundancy control to improve query representations.
Moreover, MMR markedly improves over Top-k selection on DL’19 and DL-HARD, with the largest advantage on DL-HARD (and significant gains over PLAID), indicating that redundancy control and coverage are particularly important for difficult queries. Overall, while Top-k selection is slightly better on DL’20, MMR yields a stronger overall effectiveness profile by avoiding near-duplicate expansion directions and making better use of the limited 
$f_e$ expansion slots.

\subsection{RQ6: Qualitative Analysis}\label{ssec:qualitative}

Table~\ref{tab:examples} illustrates the qualitative \x{behaviour} of PLAID-PRF expansions under \crc{a}
leave-one-out (LOO) analysis\crc{~\cite{wang2022colbert,wang2023effective}}. For each query, we first record the selected expansion \smd{vectors} produced by PLAID-PRF, then maintain the aligned WordPiece token IDs for these expansions and look them up and de-tokenise them back to their token forms. To quantify each token’s marginal utility, we perform LOO by removing one expansion \smd{vector} at a time while keeping all other settings unchanged, and report the resulting change in retrieval effectiveness \crc{in terms of MAP}.

Overall, the improved queries show that high-contribution expansions are typically domain-indicative tokens that sharpen query intent and improve early ranking. For ``what is a active margin'', tokens such as `east', `pacific', `continental', `earthquake', and `boundary' form a coherent geoscience context, which helps disambiguate ``active margin'' and promotes relevant passages. For ``definition of a sigmet'', expansions concentrate on aviation/meteorology vocabulary (e.g., `meteorological', `advisory', `safety'), consistent with the intended definition and yielding concentrated positive contributions. In contrast, ``tracheids are part of'' exhibits a mixed pattern: while several expansions are mildly helpful, a subset of generic or fragmentary tokens (e.g., `tr', `linking', `to', `\#\#a') receives negative LOO contributions, indicating that some expansion \smd{vectors} can introduce noisy matching signals that distract late-interaction alignment. This contrast motivates token-level or code-level filtering to reduce non-informative expansions and improve robustness.

\newcommand{\tok}[1]{\texttt{#1}}
\definecolor{lightred}{RGB}{255,220,220}
\definecolor{lightblue}{RGB}{220,230,255}
\newcommand{\ptok}[2]{\begingroup\setlength\fboxsep{1pt}\colorbox{lightred!#1}{\strut\texttt{#2}}\endgroup}
\newcommand{\ntok}[2]{\begingroup\setlength\fboxsep{1pt}\colorbox{lightblue!#1}{\strut\texttt{#2}}\endgroup}

\begin{table}[t]
\centering
    \caption{Qualitative examples of PLAID-PRF expansions. Token background \x{colour} indicates the leave-one-out marginal contribution on retrieval effectiveness (darker colour denotes higher contribution, red for $\Delta>0$ and blue for $\Delta<0$).}\vspace{-.5\baselineskip}
\begin{adjustbox}{max width=85mm}
\small
\setlength{\tabcolsep}{3pt}
\renewcommand{\arraystretch}{1.15}
\begin{tabular}{p{28mm}|p{75mm}}
\toprule
Original Query & Expansion Tokens \\
\midrule
what is a active margin &
\ptok{80}{east}, \ptok{50}{[CLS]}, \ptok{66}{pacific}, \ptok{78}{crash}, \ptok{54}{[SEP]},
\ptok{60}{continental}, \ptok{72}{of}, \ptok{64}{break}, \ptok{56}{scar}, \ptok{74}{earthquake},
\ptok{18}{boundary}, \ptok{46}{and}, \ptok{24}{margin}, \ptok{24}{\#\#ic} \\

definition of a sigmet &
\ptok{23}{area}, \ptok{80}{3}, \ptok{63}{safety}, \tok{a}, \ptok{60}{criteria}, \ptok{22}{advisory},
\ptok{23}{to}, \ptok{23}{\#\#ctive}, \ptok{80}{[CLS]}, \ptok{67}{of}, \ptok{16}{meteorological},
\ptok{23}{issued}, \ptok{23}{\#\#c}, \ptok{23}{(} \\

tracheids are part of &
\ptok{50}{one}, \ntok{49}{system}, \ptok{42}{of}, \ptok{44}{vessel}, \ptok{37}{x}, \ptok{31}{[CLS]},
\ptok{45}{maturity}, \ptok{47}{other}, \ptok{44}{.}, \ptok{34}{the},
\ntok{78}{tr}, \ntok{60}{linking}, \ntok{80}{to}, \ntok{71}{\#\#a} \\
\bottomrule
\end{tabular}
\end{adjustbox}
\label{tab:examples}
\end{table}

\subsection{RQ7: Out-of-domain Performance}\label{ssec:ood}

\crcBeir{\textbf{Experimental Setup.} We now examine PLAID-PRF on the out-of-domain retrieval BEIR benchmarks. In particular, as these datasets widely differ, we vary $\beta$, which controls the contribution of expansion \smd{vectors}, separately on each of the selected BEIR datasets, i.e., DBPedia, NFCorpus, TREC-COVID and Touché-2020 (v2).}

\begin{figure}
    \centering
    \includegraphics[width=\linewidth]{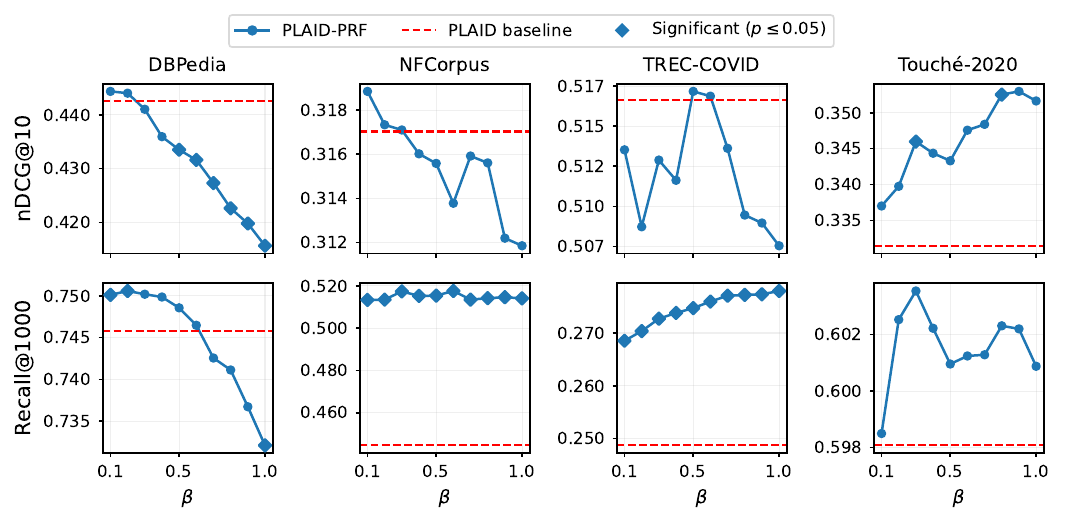}
    \vspace{-2\baselineskip}
    \caption{Effect of $\beta$ of PLAID-PRF across selected BEIR datasets, measured by nDCG@10 \& Recall@1000.}
    \label{fig:beir_curve}
\end{figure}

\noindent \crcBeir{\textbf{Results.} Figure~\ref{fig:beir_curve} shows the performance of the various $\beta$ values for nDCG@10 and Recall@1000. We observe that while for some datasets, like DBpedia and NFCorpus, PLAID-PRF does not benefit precision, and small $\beta$ values are preferred; TREC-Covid benefits for $0.5 \leq \beta \leq 0.6$; in contrast, Touché-2020 benefits for all values of $\beta$. Indeed, Touché-2020 is a counter-argument retrieval dataset, and PRF allows more relevant counter-arguments to be retrieved than the relevance trained\crcf{-}retrieval model. On the other hand, for Recall@1000, we observe that for all datasets performance is markedly enhanced across a wider range of $\beta$ values than for nDCG@10 (significantly so for 3 datasets).}

\crcBeir{Overall, in response to RQ7, these results suggest that the proposed feedback method remains effective beyond the training domain and can improve both top-ranked precision and deeper retrieval coverage in zero-shot settings.}

\section{\crcIo{Conclusions}}

In this paper, we propose \name{}, a \crcIo{novel} pseudo-relevance feedback method within the quantised late-interaction retrieval paradigm. PLAID-PRF builds the feedback model directly from PLAID/ColBERTv2 centroid codes, enabling lightweight usefulness scoring over the PRF set without operating in the full-precision space. A small set of weighted expansion \smd{vectors} is then selected (e.g., via MMR) and appended to the original query for a second-pass PLAID rerun. Experiments on \crcIo{the} TREC Deep Learning and DL-HARD \crc{datasets} show that PLAID-PRF improves nDCG@10 and early precision over PLAID while keeping latency overhead modest, yielding a competitive effectiveness–efficiency operating point among existing strong neural baselines. 
\crc{On out-of-domain BEIR benchmarks, PLAID-PRF also remains competitive and often improves upon PLAID\crcf{, particularly for Recall}.} These results indicate that PRF can be made practical for PLAID-style retrieval when feedback modelling and selection are aligned with the index's compressed structure, \crcIo{opening up the possibility of practical PRF deployments on state-of-the-art PLAID-based search engines.}

For future work, \x{a natural direction suggested by our sensitivity analysis is}
adaptive query expansion for quantised late-interaction retrieval, where the system \x{selects query-specific expansion hyperpamaters.}
For instance, expanding more aggressively for under-specified queries while keeping minimal expansions for precise navigational queries, \crcIo{to maximise potential effective improvements while avoid degradation that can hinder user experience.}

\section*{Acknowledgements}
\crc{Xiao Wang acknowledges support from the Fundamental Research Funds for the Central Universities in UIBE (Grant No.24QN01) and from the Humanities and Social Sciences Research Project of the Ministry of Education of China (Grant No.25YJCZH267).
The authors are thankful to reviewers for their constructive comments and suggestions, and Zhili Shen for insightful feedback.}

\bibliographystyle{ACM-Reference-Format}
\bibliography{reference}

\end{document}